\documentclass[12pt,smallextended,natbib,runningheads,epsfig]{article}
\usepackage{latexsym, amsfonts}
\usepackage{amssymb, amsmath}
\usepackage{amsfonts}
\usepackage{amsfonts}
\usepackage{graphicx}
\usepackage{adjustbox}
\usepackage{setspace}
\usepackage{rotating}
\usepackage{tikz}
\usepackage{bigints}
\usepackage{bm}
\usepackage{multirow}
\usepackage{changepage}  
\usepackage{subcaption}
\usepackage{hyperref}
\usepackage{color}
\usepackage{comment}

\newcommand{\be}{\begin{equation}}
\newcommand{\ee}{\end{equation}}
\newcommand{\bea}{\begin{eqnarray}}
\newcommand{\eea}{\end{eqnarray}}
\newcommand{\bean}{\begin{eqnarray*}}
\newcommand{\eean}{\end{eqnarray*}}
\newcommand{\brray}{\begin{array}}
\newcommand{\erray}{\end{array}}
\newcommand{\ben}{\begin{equation}{nonumber}}
\newcommand{\een}{\end{equation}{nonumber}}

\newtheorem{dfn}{Definition}[section]
\newtheorem{thm}[dfn]{Theorem}
\newtheorem{lmma}[dfn]{Lemma}
\newtheorem{ppsn}[dfn]{Proposition}
\newtheorem{crlre}[dfn]{Corollary}
\newtheorem{xmpl}[dfn]{Example}
\newtheorem{rmrk}[dfn]{Remark}

\newcommand{\bdfn}{\begin{dfn}}
\newcommand{\bthm}{\begin{thm}}
\newcommand{\blmma}{\begin{lmma}}
\newcommand{\bppsn}{\begin{ppsn}}
\newcommand{\bcrlre}{\begin{crlre}}
\newcommand{\bxmpl}{\begin{xmpl}}
\newcommand{\brmrk}{\begin{rmrk}}
\newcommand{\edfn}{\end{dfn}}
\newcommand{\ethm}{\end{thm}}
\newcommand{\elmma}{\end{lmma}}
\newcommand{\eppsn}{\end{ppsn}}
\newcommand{\ecrlre}{\end{crlre}}
\newcommand{\exmpl}{\end{xmpl}}
\newcommand{\ermrk}{\end{rmrk}}

\def\a*{{\cal A}_{h,*}}
\def\B{{\cal B}(h)}
\def\B1{{\cal B}_1(h)}
\def\b{{\cal B}^{\rm s.a.}(h)}
\def\b1{{\cal B}^{\rm s.a.}_1(h)}

\numberwithin{equation}{section}

\begin{document}
\begin{center}
{\Large {\bf Cluster-Based Bayesian SIRD Modeling of Chickenpox Epidemiology in India}}\\[1.5ex]
{\large Nayana Mukherjee\textsuperscript{1} and Chitradipa Chakraborty\textsuperscript{2}}\\
\end{center}

\begin{adjustwidth}{-0.5cm}{-0.5cm} 
\begin{abstract}
\footnotesize
This study presents a cluster-based Bayesian SIRD model to analyze the epidemiology of chickenpox (varicella) in India, utilizing data from 1990 to 2021. We employed an age-structured approach, dividing the population into juvenile, adult, and elderly groups, to capture the disease's transmission dynamics across diverse demographic groups. The model incorporates a Holling-type incidence function, which accounts for the saturation effect of transmission at high prevalence levels, and applies Bayesian inference to estimate key epidemiological parameters, including transmission rates, recovery rates, and mortality rates. The study further explores cluster analysis to identify regional clusters within India based on the similarities in chickenpox transmission dynamics, using criteria like incidence, prevalence, and mortality rates. We perform K-means clustering to uncover three distinct epidemiological regimes, which vary in terms of outbreak potential and age-specific dynamics. The findings highlight juveniles as the primary drivers of transmission, while the elderly face a disproportionately high mortality burden. Our results underscore the importance of age-targeted interventions and suggest that regional heterogeneity should be considered in public health strategies for disease control. The model offers a transparent, reproducible framework for understanding long-term transmission dynamics and supports evidence-based planning for chickenpox control in India.  The practical utility of the model is further validated through a simulation study.

\vspace{0.15 in}

\noindent {\bf Keywords:} Age-structured Approach; Bayesian Inference;  Chickenpox; Cluster Analysis; SIRD Model; Transmission Dynamics

\end{abstract}
\end{adjustwidth}

\renewcommand{\thefootnote}{\fnsymbol{footnote}} 

\footnotetext{\hspace{-0.5em}$^{1}$Mathematics Department, École Centrale School of Engineering, Mahindra University, Hyderabad, Telangana 500043, India. \\
\hspace{1em}Email: \texttt{nayana.mukherjee@mahindrauniversity.edu.in}} 

\footnotetext{\hspace{-0.5em}$^{2}$Beijing Key Laboratory of Topological Statistics and Applications for Complex Systems, Beijing Institute of Mathematical Sciences and Applications, Beijing 101408, China. \\
\hspace{1em}Email: \texttt{chitradipachakraborty@gmail.com} (corresponding author)}

\setcounter{footnote}{0} 
\renewcommand{\thefootnote}{\arabic{footnote}} 

\vspace{1 in}
\section{Introduction}

Infectious disease modeling has become an indispensable tool in modern epidemiology, enabling researchers to reconstruct transmission dynamics, quantify epidemiological parameters, and assess long-term disease trends \cite{anderson1992infectious, keeling2008modeling}. Mathematical models translate biological assumptions into quantitative frameworks that describe the spread of pathogens within host populations \cite{hethcote2000mathematics, a2}. Among the broad range of available approaches, compartmental models have proven especially powerful because of their clarity, interpretability, and analytical tractability. By dividing a population into discrete health states and describing the transitions between them through systems of differential equations, these models capture the essential mechanisms of infection, recovery, and immunity.

The classical Susceptible--Infectious--Recovered (SIR) model introduced by Kermack and McKendrick (1927) \cite{kermack1927contribution} remains one of the most influential frameworks in mathematical epidemiology. In the SIR model, individuals move sequentially through three epidemiological states: susceptible $(S)$, infectious $(I)$, and recovered $(R)$. The rate of change in each compartment is governed by a set of coupled nonlinear ordinary differential equations (ODEs) \cite{ionides2006inference},
\[
\frac{dS}{dt} = -\beta \frac{S I}{N}, \quad 
\frac{dI}{dt} = \beta \frac{S I}{N} - \gamma I, \quad
\frac{dR}{dt} = \gamma I,
\]
where $\beta$ is the transmission rate, $\gamma$ is the recovery rate, and $N = S + I + R$ denotes the total population size. The bilinear incidence term $\beta SI / N$ represents the rate of new infections under the assumption of homogeneous mixing, meaning that each susceptible individual has an equal probability of contact with an infectious individual. The recovery term $\gamma I$ accounts for the removal of infectious individuals from the transmission chain, either through recovery and immunity or isolation.
Despite its simplicity, the SIR model captures fundamental epidemic characteristics such as the rise and fall of infection waves and the existence of an epidemic threshold. It also provides direct insight into the balance between infection spread and recovery processes. However, the SIR model assumes that all removed individuals recover, which is unrealistic for diseases with significant mortality. For such cases, the Susceptible--Infectious--Recovered--Deceased (SIRD) model extends the SIR system by explicitly incorporating disease-induced deaths \cite{brauer2017mathematical}. In this formulation, infectious individuals may either recover at rate $\gamma$ or die from the disease at rate $\mu$, yielding the extended system
\[
\frac{dS}{dt} = -\beta \frac{S I}{N}, \quad 
\frac{dI}{dt} = \beta \frac{S I}{N} - (\gamma + \mu) I, \quad
\frac{dR}{dt} = \gamma I, \quad
\frac{dD}{dt} = \mu I.
\]
The SIRD model thus distinguishes between recovery and mortality outcomes while preserving the mass balance $N = S + I + R + D$. The parameters $\beta$, $\gamma$, and $\mu$ represent core biological processes that determine the speed and magnitude of an epidemic \cite{a3, bjornstad2018epidemics, a5}. For the SIR and SIRD frameworks with standard mass-action incidence, a key epidemiological quantity derived from these parameters is the basic reproduction number, $R_{0}= \frac{\beta}{\gamma + \mu}$.
 \cite{dietz1993estimation, diekmann1990definition, roberts2015model}, serves as a concise measure of transmission potential.
This expression reveals that $R_{0}$ is not a fixed input but rather a consequence of the underlying epidemiological rates. It depends jointly on how rapidly new infections occur (through $\beta$) and how quickly infectious individuals are removed (through $\gamma + \mu$). Conceptually, $R_{0} > 1$ indicates conditions under which an epidemic can invade, while $R_{0} < 1$ implies that infection will die out. In practice, parameter estimation focuses on $\beta$, $\gamma$, and $\mu$, and $R_{0}$ is computed subsequently from these inferred quantities \cite{van2002reproduction}. Small changes in these parameters can produce large differences in epidemic trajectories, highlighting the need for accurate estimation from data. Our modeling workflow begins with the formulation of the SIRD system using a Holling-type saturating incidence function \cite{holling1959some}. We ensure that all parameters share a consistent temporal basis and that model variables are dimensionally homogeneous. The system is numerically integrated using an ODE solver consistent with the chosen time step, and the parameters $\beta$, $\gamma$, and $\mu$ are estimated using robust optimization techniques discussed in later sections.

One of the challenges in infectious disease modeling is the heterogeneity in transmission patterns across different regions, populations, and even within specific age groups. To address this, cluster analysis has become an essential tool for identifying subgroups within a population that exhibit distinct epidemiological characteristics. Cluster analysis assumes that the observed population consists of several sub-groups, and the outcomes differ substantially from one subgroup to another. This approach is particularly useful for clustering longitudinal outcomes, where trajectories for individual subjects are examined and clustered based on these trajectories. Generally, there are two primary approaches for clustering longitudinal outcomes: model-based and algorithm-based. In a model-based approach, finite mixture models are employed, where each mixing component corresponds to a particular cluster. This approach includes group-based trajectory models \cite{Nagin1999, Nagin2010}, where subjects within a cluster follow the same trajectory. Alternatively, the latent class mixed effect model \cite{Muthén1999, ProustLima2017} assumes that subject-specific trajectories vary around a common mean trajectory. On the other hand, the algorithm-based approach relies on computational algorithms for clustering without assuming a probability distribution for the dataset. Methods like $k$-means clustering \cite{Genolini2016}, hierarchical clustering \cite{Zhou2023}, and correlation-based clustering \cite{PintoCosta2023} fall under this category.
While these methods are widely used for clustering univariate longitudinal outcomes, recent work has proposed Bayesian latent-class models for multivariate longitudinal data \cite{Chakraborty2025, Chattopadhyay2025, Kundu2025}. By employing cluster analysis alongside the SIRD model, this study aims to identify regional clusters within India that exhibit similar chickenpox transmission dynamics. This approach provides valuable insights into the age-specific and regional variations in disease burden, supporting the development of targeted public health interventions. The proposed model is estimated within a Bayesian framework using Markov Chain Monte Carlo (MCMC) methods implemented in JAGS \cite{plummer2003}.

Our work in this paper is motivated by the availability of long-term chickenpox (varicella) data in India, covering the period 1990–2021. Chickenpox, caused by the varicella-zoster virus, is a highly contagious disease that primarily affects children but can lead to more severe outcomes among adults and older individuals, making it a suitable case study for age-structured transmission analysis \cite{vazquez2004epidemiology}. Recent studies have shown significant regional variability in chickenpox transmission within India \cite{singh2011, vaidya2018}. Age-specific factors also significantly impact the epidemiology of chickenpox, as adults and the elderly experience higher mortality rates compared to children \cite{mishra2021}. The main contribution of this paper is the integration of clustering with an age-structured SIRD model to characterize regional heterogeneity in chickenpox transmission dynamics across India. Unlike the previous studies that either analyze regions independently or assume homogeneous transmission parameters, we jointly model age-specific SIRD dynamics while simultaneously clustering Indian states and union territories into subgroups with similar epidemiological patterns. This approach enables data-driven identification of regions with comparable transmission, recovery, and mortality profiles, while accounting for temporal dependence and age stratification. 

The remainder of the paper is organized as follows. Section~2 describes the proposed modeling framework, including the formulation of the age-structured SIRD model and the clustering approach used to capture regional heterogeneity in transmission dynamics. Section~3 presents the empirical analysis of the Chickenpox dataset, and interprets the findings. Section 4 summarizes the findings of a simulation study. Some concluding remarks are given in Section 5. 

\section{Model and Method}

In this study, we develop a modeling framework using annual chickenpox data from the Global Burden of Disease (GBD) dataset spanning the years 1990 to 2021. The dataset covers 28 Indian states and union territories, with Delhi treated separately due to its distinct demographic and epidemiological characteristics. The dataset provides information on incidence, prevalence, and deaths stratified by three age groups: Juvenile (0--17 years), Adult (18--59 years), and Old (60+ years). For modeling purposes, the population was scaled to a standard reference of $P = 100{,}000$. 
\subsection{Age-Structured SIRD Model with Holling-Type Incidence}

In this age-structured SIRD model, individuals in each age group are classified into four health states: susceptible ($S$), infected ($I$), recovered ($R$), and dead ($D$). A susceptible individual may become infected after contact with an infected individual. The probability of such transmission depends on how frequently the different age groups interact with each other, how many infected individuals are currently present, and a saturation effect which ensures that the risk of infection does not grow indefinitely as the number of infected individuals increases. This saturation is modeled using a Holling-type functional form. Once infected, individuals may either recover or die from the disease. 

To capture the epidemiological transmission dynamics of chickenpox, we employed a discrete-time, age-structured SIRD (Susceptible--Infected--Recovered--Dead) model. The population is divided into 
three groups: juveniles (\(j\)), adults (\(a\)), and the old (\(o\)). Each group 
is further subdivided into four compartments:
\[
S_k(t) \quad \text{(susceptible)}, \quad
I_k(t) \quad \text{(infectious)}, \quad
R_k(t) \quad \text{(recovered)}, \quad
D_k(t) \quad \text{(disease deaths)},
\]
where \(k \in \{j,a,o\}\). The living population of each age group is $N_k(t) = S_k(t) + I_k(t) + R_k(t)$.

Transmission between age groups is governed by a Holling-type incidence function 
which accounts for saturation of contacts at high prevalence. The force of 
infection experienced by group \(k\) is defined as
\begin{equation*}
\lambda_k(t) = \sum_{k' \in \{j,a,o\}} 
\beta_{kk'}\,\frac{I_{k'}(t)/N_{k'}(t)}{1+\alpha_{kk'}\,I_{k'}(t)/N_{k'}(t)},
\end{equation*}
where \(\beta_{kk'}\) measures the transmission intensity from infectious 
individuals in group \(k'\) to susceptibles in group \(k\), and 
\(\alpha_{kk'} \geq 0\) quantifies the degree of saturation.

\subsection{Governing Equations}
The model is described by the following system of ordinary differential 
equations:
\begin{align*}
\frac{dS_k}{dt} &= -\lambda_k(t)\,S_k(t), \\[6pt]
\frac{dI_k}{dt} &= \lambda_k(t)\,S_k(t) - (\gamma_k+\mu_k)\,I_k(t), \\[6pt]
\frac{dR_k}{dt} &= \gamma_k\,I_k(t), \\[6pt]
\frac{dD_k}{dt} &= \mu_k\,I_k(t),
\end{align*}
for each \(k \in \{j,a,o\}\). The parameters \(\gamma_k\) and \(\mu_k\) denote 
the recovery rate and disease-induced mortality rate for age group \(k\), 
respectively. 

Let $s_{m,k,t}$, $i_{m,k,t}$, $r_{m,k,t}$, and $d_{m,k,t}$ denote the fraction of the population in the susceptible, infected, recovered, and cumulative death compartments, respectively, for the state $m$ and age group $k$ at year $t$. The model was initialized with
$
s_{m,k,1} = 0.97, \quad i_{m,k,1} = 0.02, \quad r_{m,k,1} = 0.01, \quad d_{m,k,1} = 0.
$ The compartmental dynamics were defined by $
\text{new\_inf}_{m,k,t} = \beta_{m,k} s_{m,k,t} i_{m,k,t}, 
\text{new\_rec}_{m,k,t} = \gamma_{m,k} i_{m,k,t}, 
\text{new\_death}_{m,k,t} = \mu_{m,k} i_{m,k,t},
$
with updates
\[
\begin{aligned}
s_{m,k,t+1} &= \max(s_{m,k,t} - \text{new\_inf}_{m,k,t}, 0), \\
i_{m,k,t+1} &= \max(i_{m,k,t} + \text{new\_inf}_{m,k,t} - \text{new\_rec}_{m,k,t} - \text{new\_death}_{m,k,t}, 0), \\
r_{m,k,t+1} &= r_{m,k,t} + \text{new\_rec}_{m,k,t}, \\
d_{m,k,t+1} &= d_{m,k,t} + \text{new\_death}_{m,k,t}.
\end{aligned}
\]

\noindent Here, $\beta_{m,k}$ represents the transmission rate, $\gamma_{m,k}$ the recovery rate, and $\mu_{m,k}$ the mortality rate for each state and age group. The basic reproduction number for k-th age group was calculated as 
\[
R_{0,k} = \frac{\beta_k}{\gamma_k + \mu_k}.
\]

\subsection{Bayesian Computation}
A Bayesian framework was adopted to estimate the model parameters. The prior distributions for the age and state-specific parameters were specified as
$
\beta_k \sim \text{Gamma}(2,0.2),  \gamma_k \sim \text{Gamma}(2,1), \text{and }  \mu_k \sim \text{Gamma}(1,100).
$
To account for underreporting and measurement errors, we introduced additional reporting factors with priors $\rho_\text{inc} \sim \text{Uniform}(0,5)$, $\rho_\text{prev} \sim \text{Uniform}(0,5)$), $\rho_\text{death} \sim \text{Uniform}(0.5, 2)$, and a dispersion parameter $\sigma_\text{prev} \sim \text{Uniform}(0.001,0.5)$.
The SIRD compartments were linked to the observed data through the following likelihoods:
\[
\begin{aligned}
y^\text{inc}_{m,k,t} &\sim \text{Poisson}(\rho_\text{inc} \text{new\_inf}_{m,k,t}), \\
y^\text{prev}_{m,k,t} &\sim \text{Normal}(\rho_\text{prev} i_{m,k,t}, \sigma_\text{prev}^2), \\
y^\text{death}_{m,k,t} &\sim \text{Poisson}(\rho_\text{death} \text{new\_death}_{m,k,t}).
\end{aligned}
\]
This joint modeling of incidence, prevalence, and mortality enhances inference on transmission dynamics. Incidence informs the rate of new infections, prevalence reflects the current infected fraction, and death data provide information on infection-related mortality. This integration allows leveraging both flow (incidence) and stock (prevalence, deaths) data while accounting for measurement error and reporting bias. 
We estimate the model parameters using RJAGS by the respective posterior medians, and compute the 95\% credible intervals based on the sample quantiles. The convergence of the chains is assessed by the trace plots and by computing the scale reduction factors \cite{gelman1992}.

\subsection{K-means Clustering Method}
\begin{figure}[h!]
    \centering
    \begin{minipage}{0.3\textwidth}
        \centering
        \includegraphics[width=\textwidth,height=3.2cm,keepaspectratio]{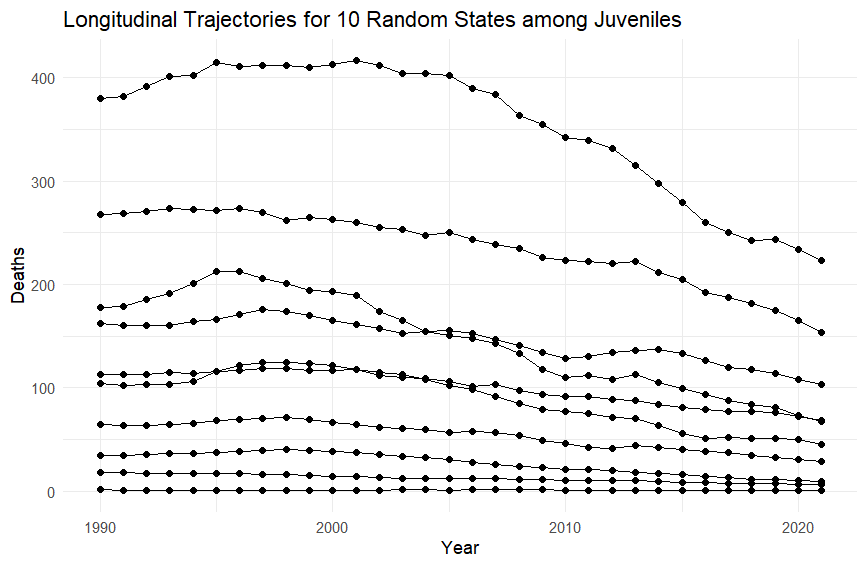}
        \subcaption{}
    \end{minipage}
    \hfill
    \begin{minipage}{0.3\textwidth}
        \centering
        \includegraphics[width=\textwidth,height=3.2cm,keepaspectratio]{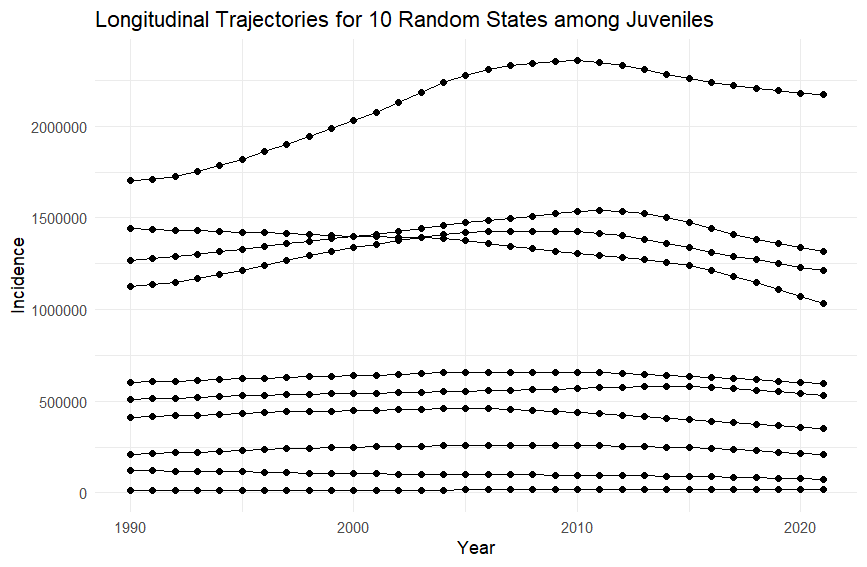}
        \subcaption{}
    \end{minipage}
    \begin{minipage}{0.3\textwidth}
        \centering
        \includegraphics[width=\textwidth,height=3.2cm,keepaspectratio]{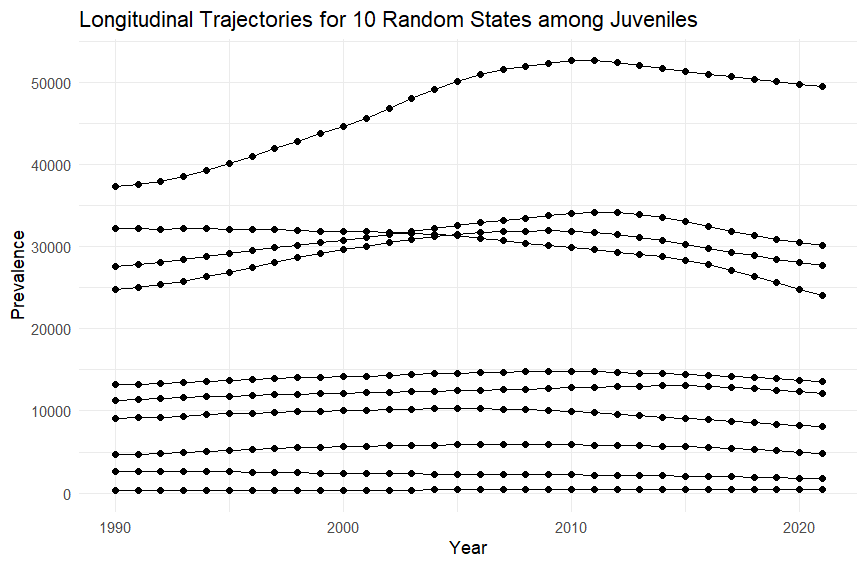}
        \subcaption{}
    \end{minipage}

    \vspace{0.3cm}

    \begin{minipage}{0.3\textwidth}
        \centering
        \includegraphics[width=\textwidth,height=3.2cm,keepaspectratio]{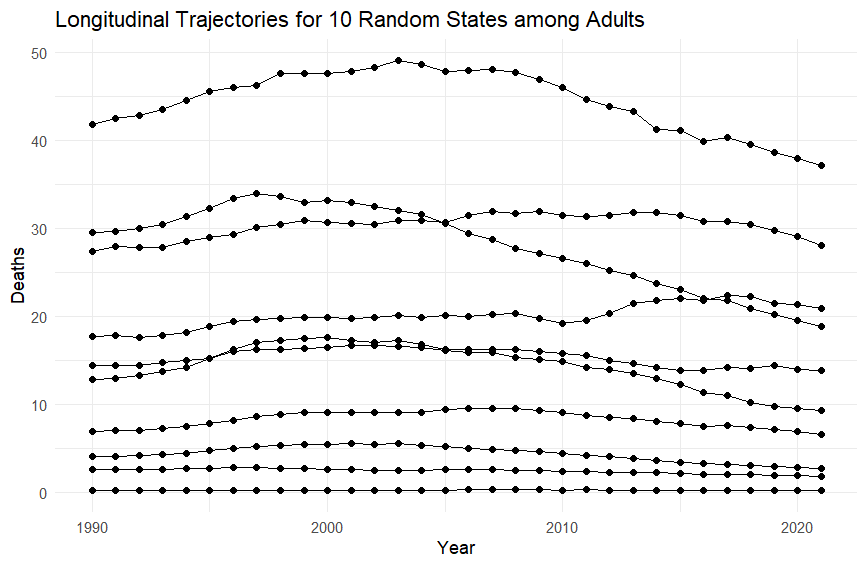}
        \subcaption{}
    \end{minipage}
    \hfill
    \begin{minipage}{0.3\textwidth}
        \centering
        \includegraphics[width=\textwidth,height=3.2cm,keepaspectratio]{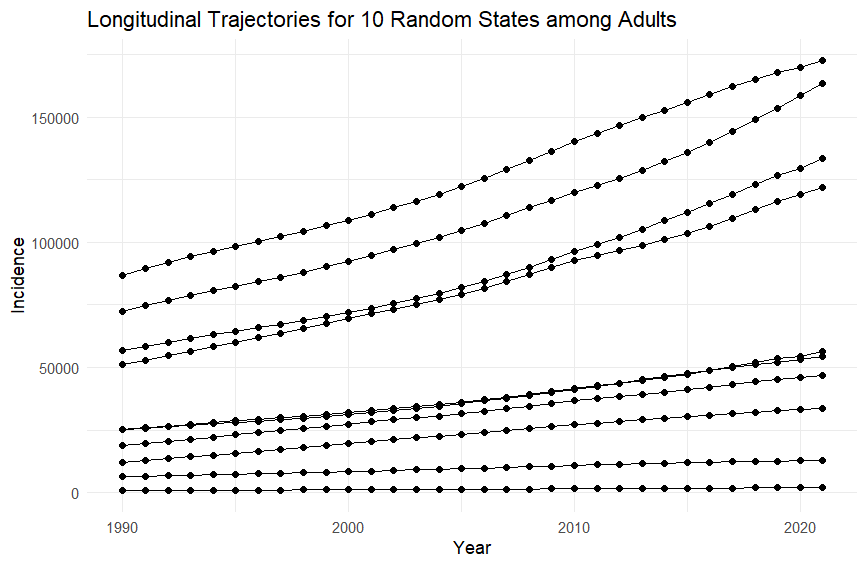}
        \subcaption{}
    \end{minipage}
    \begin{minipage}{0.3\textwidth}
        \centering
        \includegraphics[width=\textwidth,height=3.2cm,keepaspectratio]{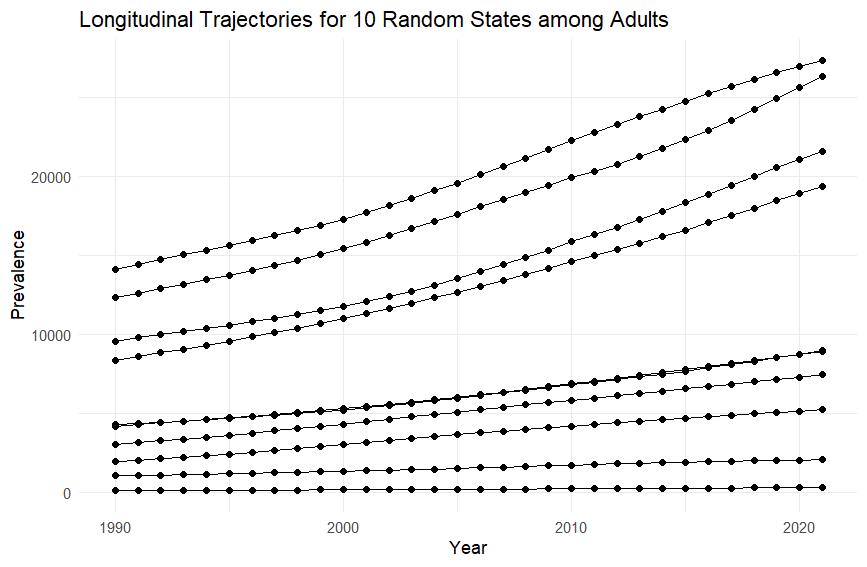}
        \subcaption{}
    \end{minipage}

    \vspace{0.3cm}

    \begin{minipage}{0.3\textwidth}
        \centering
        \includegraphics[width=\textwidth,height=3.2cm,keepaspectratio]{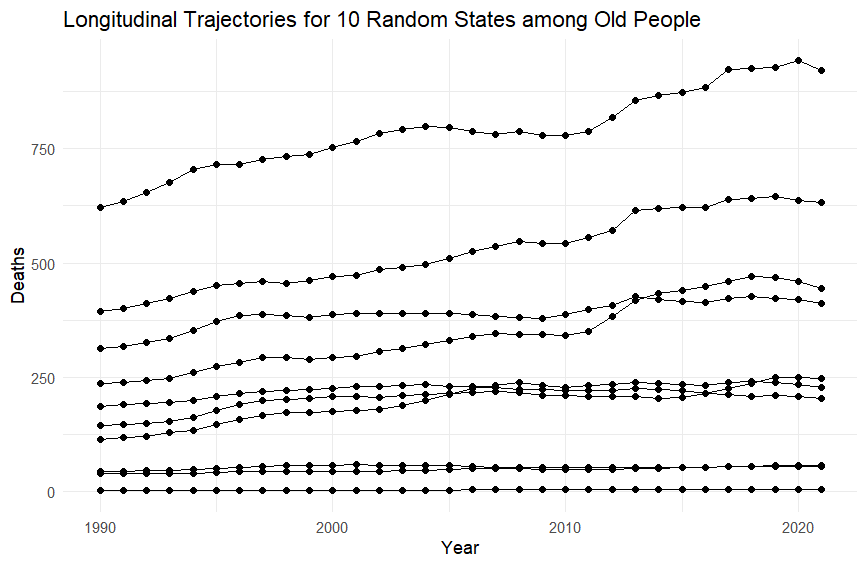}
        \subcaption{}
    \end{minipage}
    \hfill
    \begin{minipage}{0.3\textwidth}
        \centering
        \includegraphics[width=\textwidth,height=3.2cm,keepaspectratio]{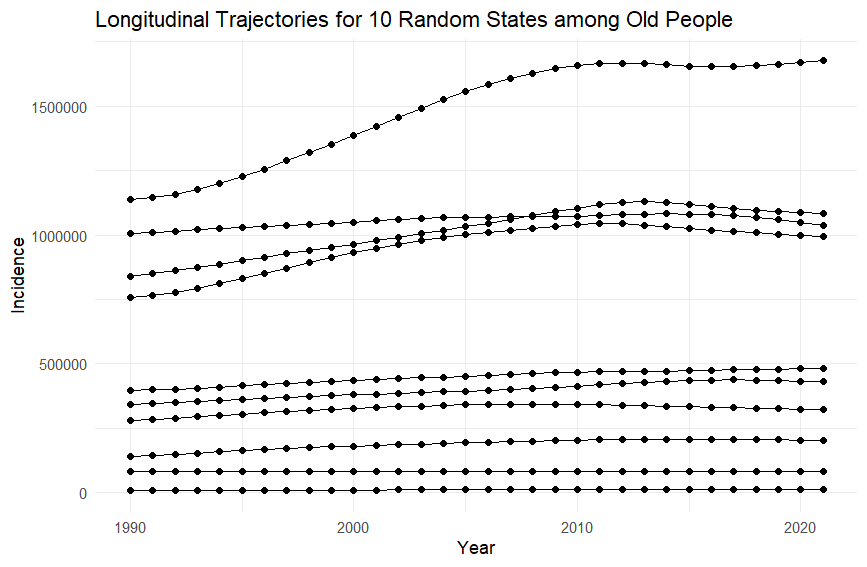}
        \subcaption{}
    \end{minipage}
    \begin{minipage}{0.3\textwidth}
        \centering
        \includegraphics[width=\textwidth,height=3.2cm,keepaspectratio]{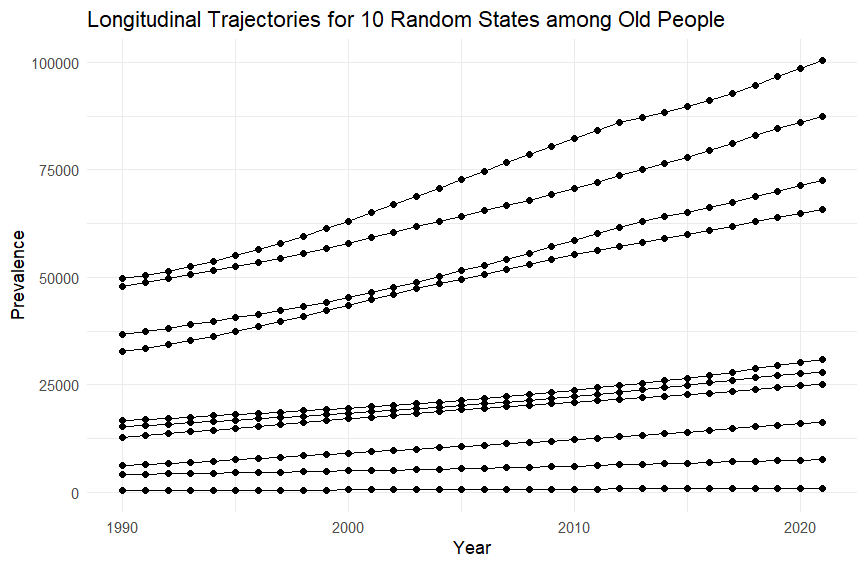}
        \subcaption{}
    \end{minipage}

    \caption{Mean longitudinal trajectories of three features for 10 randomly selected states in the Chickenpox dataset. (a--c) Deaths, incidence, and prevalence among juveniles; (d--f) deaths, incidence, and prevalence among adults; (g--i) deaths, incidence, and prevalence among the elderly.}
    \label{fig:long}
\end{figure}

Figure~\ref{fig:long} displays the longitudinal trajectories of Deaths, Incidence, and Prevalence for juveniles, adults, and older individuals across ten randomly selected states. The three age groups show clearly different temporal patterns, and substantial variation is also
observed across states within each group. This heterogeneity indicates that the disease burden does not evolve uniformly across the states. The diverse trends in the data indicate the presence of underlying subgroups with distinct epidemiological behaviors, which motivates the use of clustering to uncover these hidden patterns. This approach helps to better understand the age-specific and state-specific dynamics in the chickenpox data. To capture heterogeneity\\

\noindent in epidemic dynamics across states, we applied K-means clustering to group states based on epidemiological features such as incidence, prevalence, and mortality rates over time. This clustering approach allows for the identification of subgroups of states with similar transmission characteristics and disease progression patterns. States within the same cluster share comparable SIRD dynamics, reflecting homogeneous epidemiological behavior within each group. The longitudinal trajectories (Figure~\ref{fig:long}) are presented for exploratory purposes to illustrate temporal heterogeneity across states and age groups and to motivate the clustering analysis. The optimal number of clusters (K) was selected using multiple criteria to ensure robustness and avoid overfitting. We considered the following methods:

\noindent \textbf{Elbow Method:} The within-cluster sum of squares (WCSS)  \cite{thorndike1953} is plotted against the number of clusters to identify the point where the rate of decrease in WCSS slowed significantly, which suggested a reasonable choice of clusters.

\noindent \textbf{Silhouette Score:} The silhouette score \cite{rousseeuw1987} is calculated for each value of K to assess the quality of clustering. A higher silhouette score indicates better-defined and more distinct clusters.

\noindent \textbf{AIC and BIC:} These information criteria are computed for each K to balance model fit with complexity. Lower values of AIC \cite{akaike1974} and BIC \cite{schwarz1978} suggest better models with fewer parameters.

\noindent The resulting clustering approach has been used to reveal distinct regional patterns in chickenpox transmission and recovery, highlighting meaningful epidemiological similarities among states within the same cluster.

\section{Data Analysis and Results}





\begin{figure}[htbp]
    \centering
    \begin{minipage}{0.45\textwidth}
        \centering
        \includegraphics[height=4.4cm]{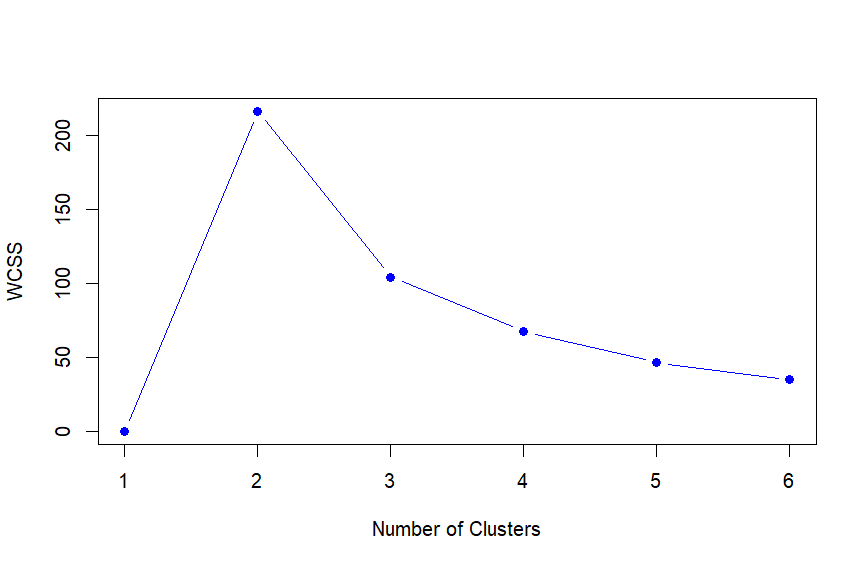}
        \subcaption{}
        \label{fig:wcss}
    \end{minipage}
    \hfill
    \begin{minipage}{0.45\textwidth}
        \centering
        \includegraphics[height=4.4cm]{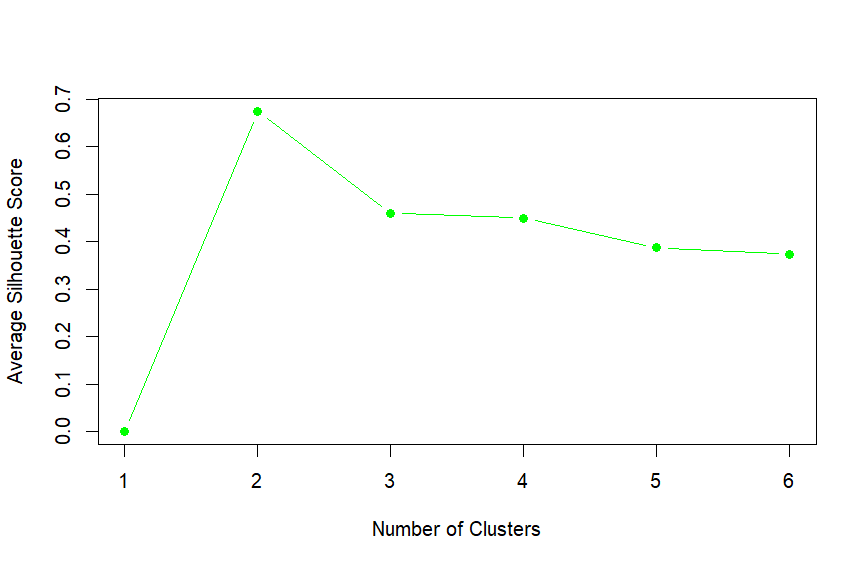}
        \subcaption{}
        \label{fig:s_score}
    \end{minipage}
 \caption{(a) Elbow Method and (b) Silhouette Score }   
\end{figure}

Based on the Elbow Method (Figure \ref{fig:wcss}), the WCSS showed a significant drop between K = 2 and K = 3, and the rate of decrease slowed after K = 3, suggesting that additional clusters beyond 3 would not provide substantial improvement in clustering. The Silhouette Score also suggested that the clustering quality improved with K = 3 (Figure \ref{fig:s_score}), indicating that the clusters were better separated and more cohesive compared to K = 2. While AIC and BIC favored a simpler model with K = 2 (Table~\ref{tab:optK}), the overall clustering quality and interpretability were superior for K = 3, and therefore, we selected K = 3 as the optimal number of clusters.

The three distinct clusters of Indian states are based on chickenpox epidemiological patterns: Cluster 1 (19 states, 63.4\%) includes Andhra Pradesh, Assam, Bihar, Chhattisgarh, Delhi, Gujarat, Haryana, Jharkhand, Karnataka, Kerala, Madhya Pradesh, Maharashtra, Odisha, Punjab, Rajasthan, Tamil Nadu, Telangana, Uttarakhand, and West Bengal; 
Clus-

\begin{table}[h!]
\centering
\begin{tabular}{|c|c|c|c|}
\hline
\textbf{$\boldsymbol{K}$} & \textbf{Silhouette Scores} & \textbf{AIC} & \textbf{BIC} \\
\hline
2 & 0.6744202 & 109.2882 & 144.3182\\
\hline
3 & 0.4594910 & 113.4381 & 166.6836\\
\hline
4 &  0.4494454 & 126.4697 & 197.9308\\
\hline
5 & 0.3872293  & 141.2778 & 230.9544\\
\hline
6 & 0.3730322 & 158.6279 & 266.5201\\
\hline
\end{tabular}
\caption{Criteria for Optimum K \label{tab:optK}}
\end{table}

\noindent ter 2 (Uttar Pradesh, 3.3\%) contains only Uttar Pradesh, and 
Cluster 3 (10 states, 33.3\%) comprises Arunachal Pradesh, Goa, Himachal Pradesh, Manipur, Meghalaya, Mizoram, Nagaland, Sikkim, Tripura, and Union Territories other than Delhi. 

\subsection{Cluster-specific Findings}

\begin{figure}[h!]
  \centering

  \begin{subfigure}{\linewidth}
    \centering
    \includegraphics[width=\linewidth, height=4cm]{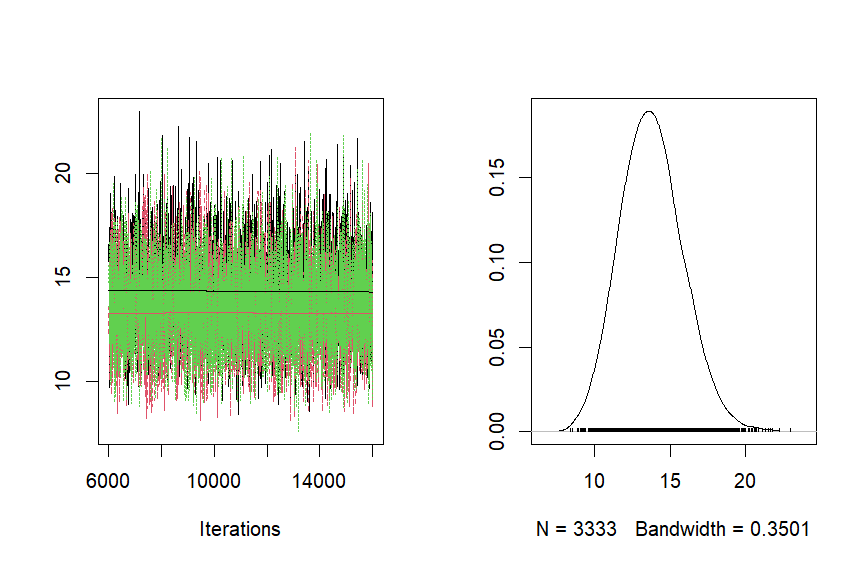}
    \subcaption{}
  \end{subfigure}

  \vspace{0.1cm}

  \begin{subfigure}{\linewidth}
    \centering
    \includegraphics[width=\linewidth, height=4cm]{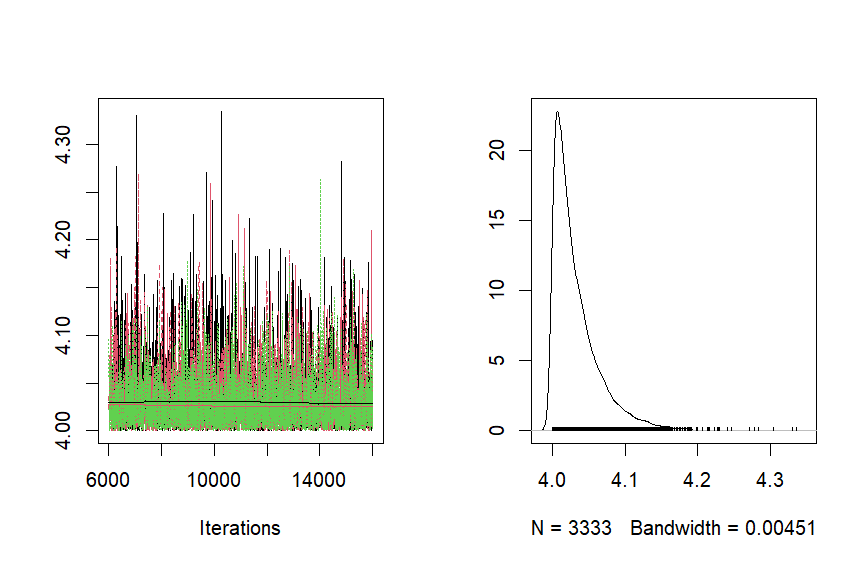}
    \subcaption{}
  \end{subfigure}

  \vspace{0.1cm}

  \begin{subfigure}{\linewidth}
    \centering
    \includegraphics[width=\linewidth, height=4cm]{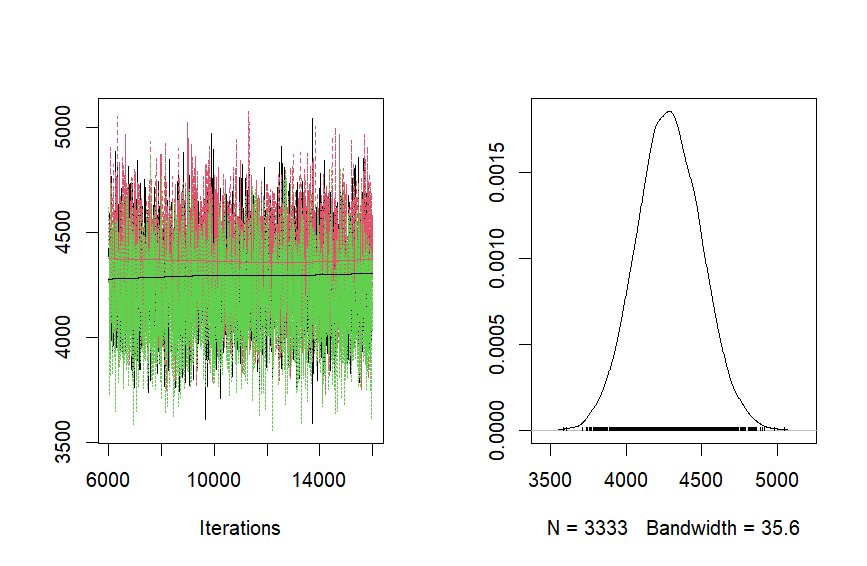}
    \subcaption{}
  \end{subfigure}

  \caption{Trace and posterior density plots for the precision parameter $\tau_{\text{prev}}$ across three MCMC chains, shown separately for (a) Cluster~1, (b) Cluster~2, and (c) Cluster~3.}
  \label{fig:tau_trace}
\end{figure}

Posterior summaries were obtained via MCMC using RJAGS with three chains, 1,000 adaptation iterations, 2,000 burn-in iterations, and 10,000 sampling iterations with a thinning factor of 3. Figure \ref{fig:tau_trace}  presents trace and posterior density plots for the precision parameter   

\noindent (inverse of the dispersion parameter) $\tau_{\text{prev}}$ based on three clusters, evaluated across three MCMC chains. The trace plots demonstrate good mixing, with the chains fluctuating around stable means and exhibiting no signs of non-convergence such as persistent trends or high autocorrelation \cite{gelman2013}. The density plots are smooth and unimodal, further supporting that the posterior distributions are well explored and stably estimated. The Gelman-Rubin convergence diagnostics, i.e., the scale reduction factors ($\hat{R}$) for the parameter, are 1.03 for Cluster 1, 1.045 for Cluster 2, and 1.004 for Cluster 3, all of which are well below the standard convergence threshold of 1.1 \cite{gelman1992}, indicating strong convergence across all clusters. These results indicate that the chains have converged adequately and that the posterior estimates are robust and reliable.

Posterior mean estimates of the transmission rate ($\beta$), recovery rate ($\gamma$), mortality rate ($\mu$), and basic reproduction number ($R_0$) varied markedly across clusters and age groups (Table~\ref{tab:cluster_summary}).\\

\begin{table}[htbp]
\centering
\begin{tabular}{lcccccc}
\hline
\textbf{Cluster} & \textbf{Age Group} & \textbf{$\beta$} & \textbf{$\gamma$} & \textbf{$\mu$} & \textbf{$R_0$} \\
\hline
Cluster 1 & Adult (18--59) & 6.403 & 5.553 & 0.114 & 1.26 \\
 & Juvenile (0--17) & 7.464 & 5.644 & 0.463 & 1.88 \\
 & Old (60+) & 5.575 & 3.141 & 0.831 & 1.31 \\
\hline
Cluster 2 & Adult (18--59) & 2.671 & 0.596 & 0.0063 & 5.42 \\
 & Juvenile (0--17) & 13.030 & 9.218 & 1.569 & 1.22 \\
 & Old (60+) & 13.001 & 8.055 & 2.657 & 1.29 \\
\hline
Cluster 3 & Adult (18--59) & 4.570 & 2.890 & 0.034 & 1.55 \\
 & Juvenile (0--17) & 8.918 & 7.669 & 0.505 & 1.1 \\
 & Old (60+) & 6.650 & 5.550 & 0.090 & 1.31 \\
\hline
\end{tabular}
\caption{Posterior mean SIRD parameters for chickenpox across age groups and clusters.}
\label{tab:cluster_summary}
\end{table}


Cluster 1 states, representing major and highly populated regions, exhibited moderate transmission rates among adults ($\beta$ = 6.40), higher transmission among juveniles ($\beta$ = 7.46) compared to adults and older adults ($\beta$ = 5.58). Recovery rates were highest in juveniles ($\gamma$ = 5.64) and adults ($\gamma$ = 5.55), but lower in older adults ($\gamma$ = 3.14). Mortality rates were lowest for adults ($\mu$ = 0.114), moderate for juveniles ($\mu$ = 0.463), and highest for older adults ($\mu$ = 0.831). These dynamics resulted in basic reproduction numbers of 1.26 (Adult), 1.88 (Juvenile), and 1.31 (Old), indicating that juveniles are the primary drivers of transmission, while older adults are at greater risk due to slower recovery and higher mortality.

Cluster 2, representing a single state or region with extreme age-specific dynamics, showed moderate transmission in adults ($\beta$ = 2.67), with recovery ($\gamma$ = 0.6) and mortality ($\mu$ = 0.0063) rates resulting in $R_0$ = 5.42. Transmission was very high among juveniles ($\beta$ = 13.03) and older adults ($\beta$ = 13.00), with moderate recovery rates ($\gamma$ = 9.22 and 8.06, respectively) and elevated mortality ($\mu$ = 1.57 and 2.66, respectively). The corresponding $R_0$ values for juveniles, and older adults were 1.22, and 1.29, respectively. These estimates highlight strong age-specific heterogeneity, with adults showing moderately high outbreak potential despite lower transmission compared to juveniles and older adults. 

Cluster 3, consisting of smaller or northeastern states, also demonstrated higher transmission among juveniles ($\beta$ = 8.92), followed by older adults ($\beta$ = 6.65) and adults ($\beta$ = 4.57). Recovery rates were highest in juveniles ($\gamma$ = 7.67) and older adults ($\gamma$ = 5.55), but moderate in adults ($\gamma$ = 2.89). Mortality was generally low across age groups ($\mu$ = 0.034–0.505), leading to $R_0$ values of 1.55 (Adult), 1.1 (Juvenile), and 1.31 (Old). This suggests lower outbreak potential in these smaller states while still emphasizing juveniles as key contributors to transmission.  

Overall, Cluster 1 states, which include highly populated regions, show higher outbreak potential among juveniles and older adults due to slower recovery and higher mortality. Cluster 2 shows strong age-specific differences, with adults exhibiting moderate outbreak potential ($R_0$ = 5.42), while juveniles and older adults maintain very high transmission. Cluster 3 states have lower transmission potential across all age groups. These findings suggest that interventions in Cluster 1 may require age-targeted strategies, such as vaccination campaigns, particularly for juveniles and the elderly, while Cluster 3 may benefit primarily from general surveillance and monitoring. The age-stratified analysis consistently shows that juveniles drive transmission, adults maintain moderate transmission with low mortality, and older adults are vulnerable due to slower recovery and higher mortality. 

\subsection{Interpretation of the Cluster-based SIRD Dynamics}

The posterior SIRD parameter estimates identify three distinct epidemic regimes across the clusters, reflecting pronounced heterogeneity in age-specific transmission dynamics. Although the governing equations are identical, variations in the transmission rate $\beta$, recovery rate $\gamma$, mortality rate $\mu$, and the strength of the saturating incidence function give rise to markedly different epidemic behaviors.

\begin{figure}[h!]
  \centering

  \begin{minipage}[b]{0.47\textwidth}
    \centering
    \includegraphics[width=\textwidth]{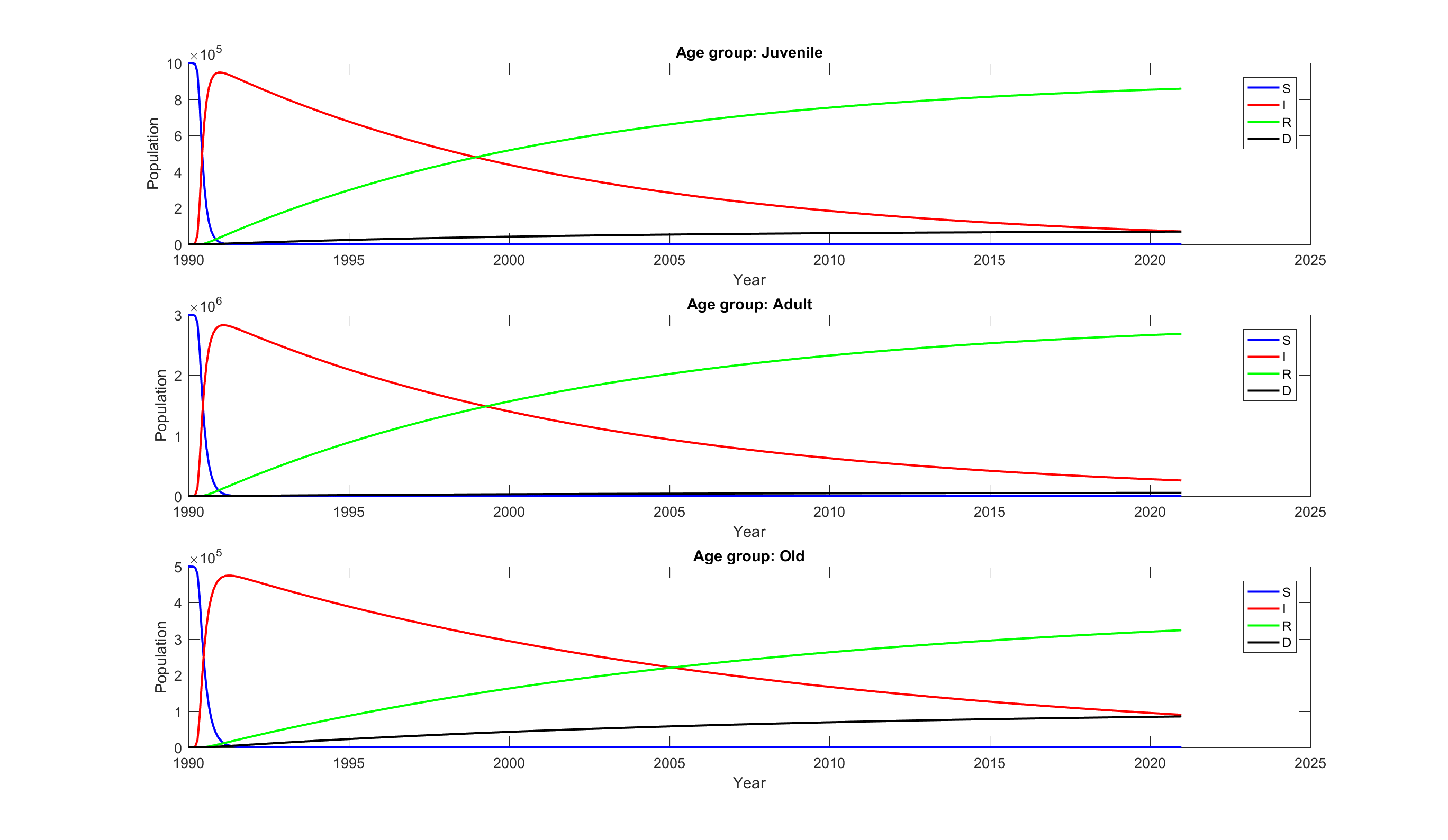}
    \subcaption{}
    \label{fig:sub1}
  \end{minipage}
  \hfill
  \begin{minipage}[b]{0.47\textwidth}
    \centering
    \includegraphics[width=\textwidth]{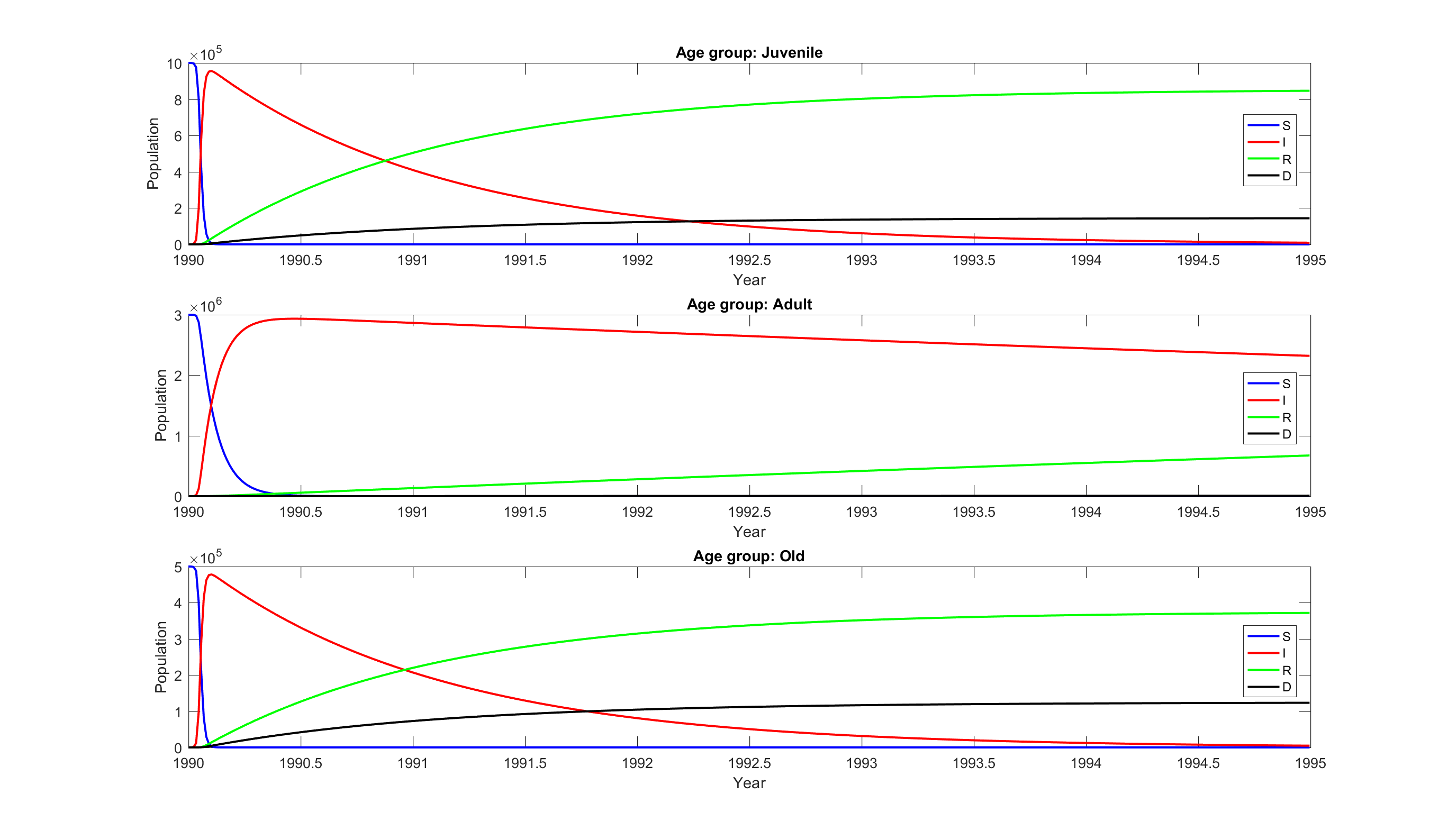}
    \subcaption{}
    \label{fig:sub2}
  \end{minipage}
  \hfill
  \begin{minipage}[b]{0.5\textwidth}
    \centering
    \includegraphics[width=\textwidth]{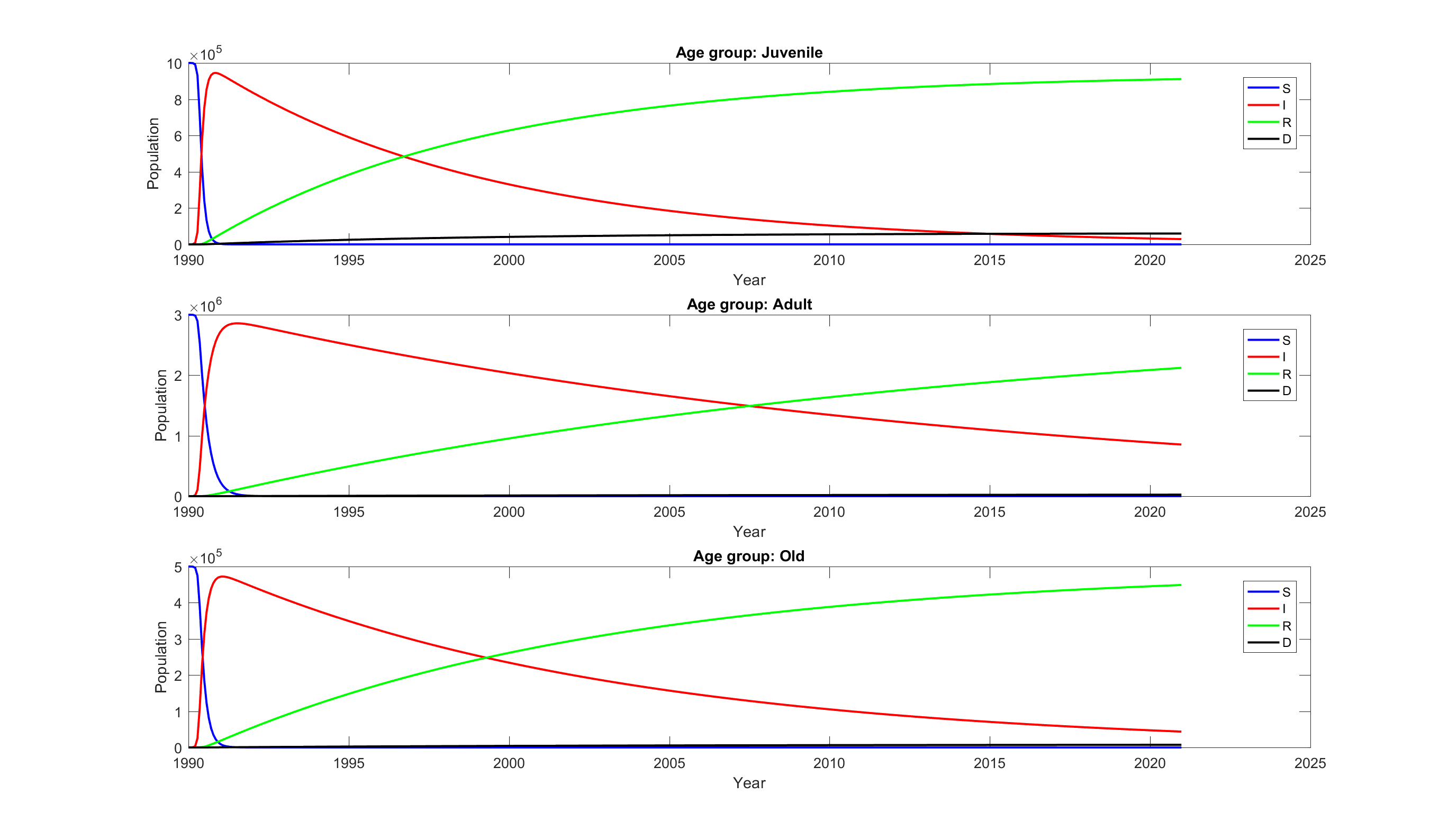}
    \subcaption{}
    \label{fig:sub3}
  \end{minipage}

  \caption{SIRD trajectories for juveniles, adults, and old-age individuals under a moderate-transmission regime using parameter values from Table \ref{tab:cluster_summary} based on three clusters; (a) Cluster 1, (b) Cluster 2, and (c) Cluster 3.}
  \label{fig:trajectories}
\end{figure}

Cluster~1 corresponds to a moderate-transmission regime with balanced transmission, recovery, and mortality rates. As shown in Figure~\ref{fig:trajectories}(a), infections rise rapidly at early times and then decay smoothly across all age groups, followed by a gradual recovery phase. The susceptible population declines steadily without an abrupt collapse, while mortality accumulates progressively, remaining moderate in juveniles and adults and higher among the elderly. The presence of saturation effects limits transmission at high prevalence, resulting in a relatively controlled epidemic evolution with prolonged transient dynamics.

\begin{figure}[h!]
    \centering
    \begin{minipage}{0.48\textwidth}
        \centering
        \includegraphics[width=\textwidth,height=6.5cm]{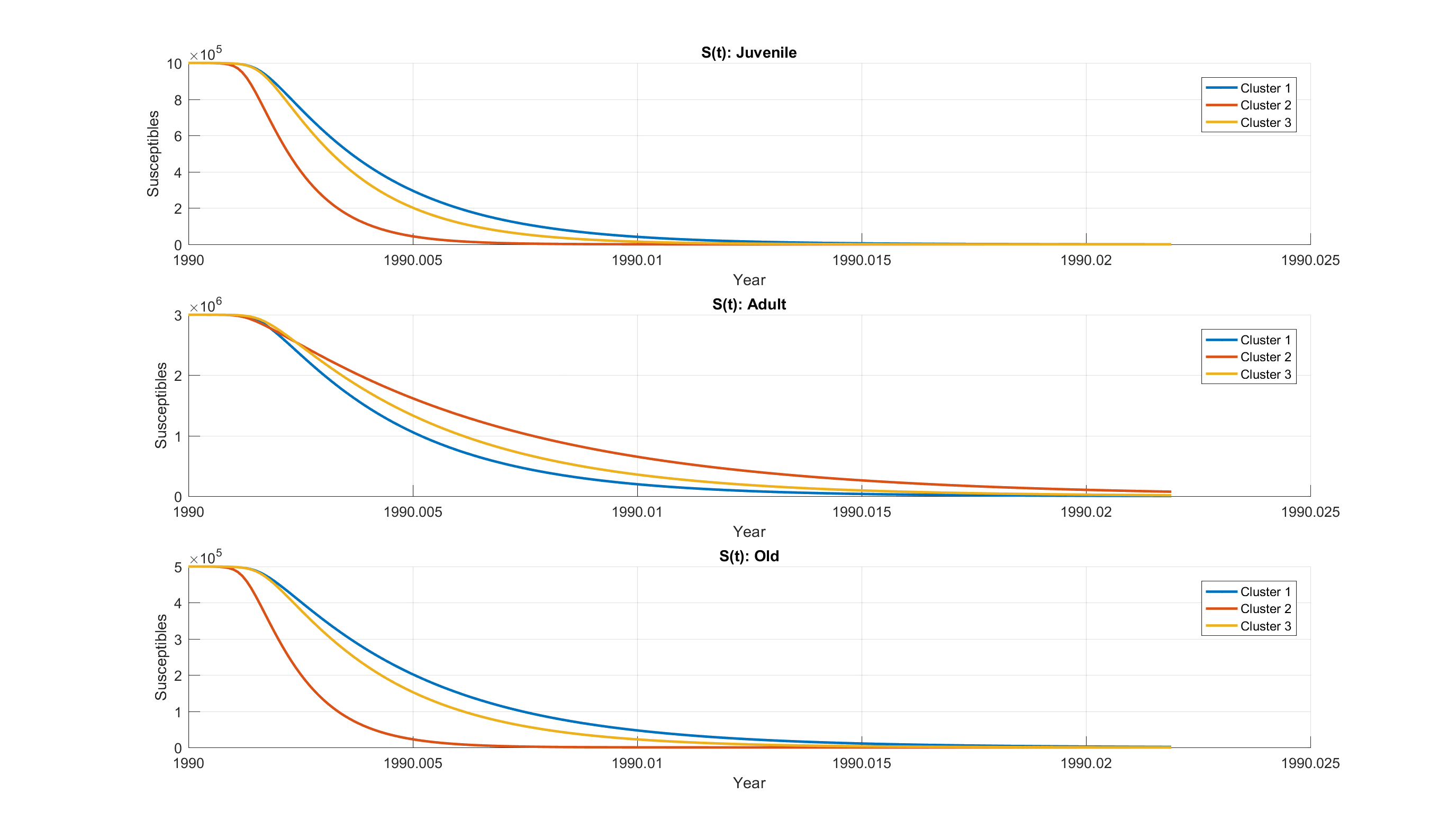}
        \subcaption{}
    \end{minipage}
    \hfill
    \begin{minipage}{0.48\textwidth}
        \centering
        \includegraphics[width=\textwidth,height=6.5cm]{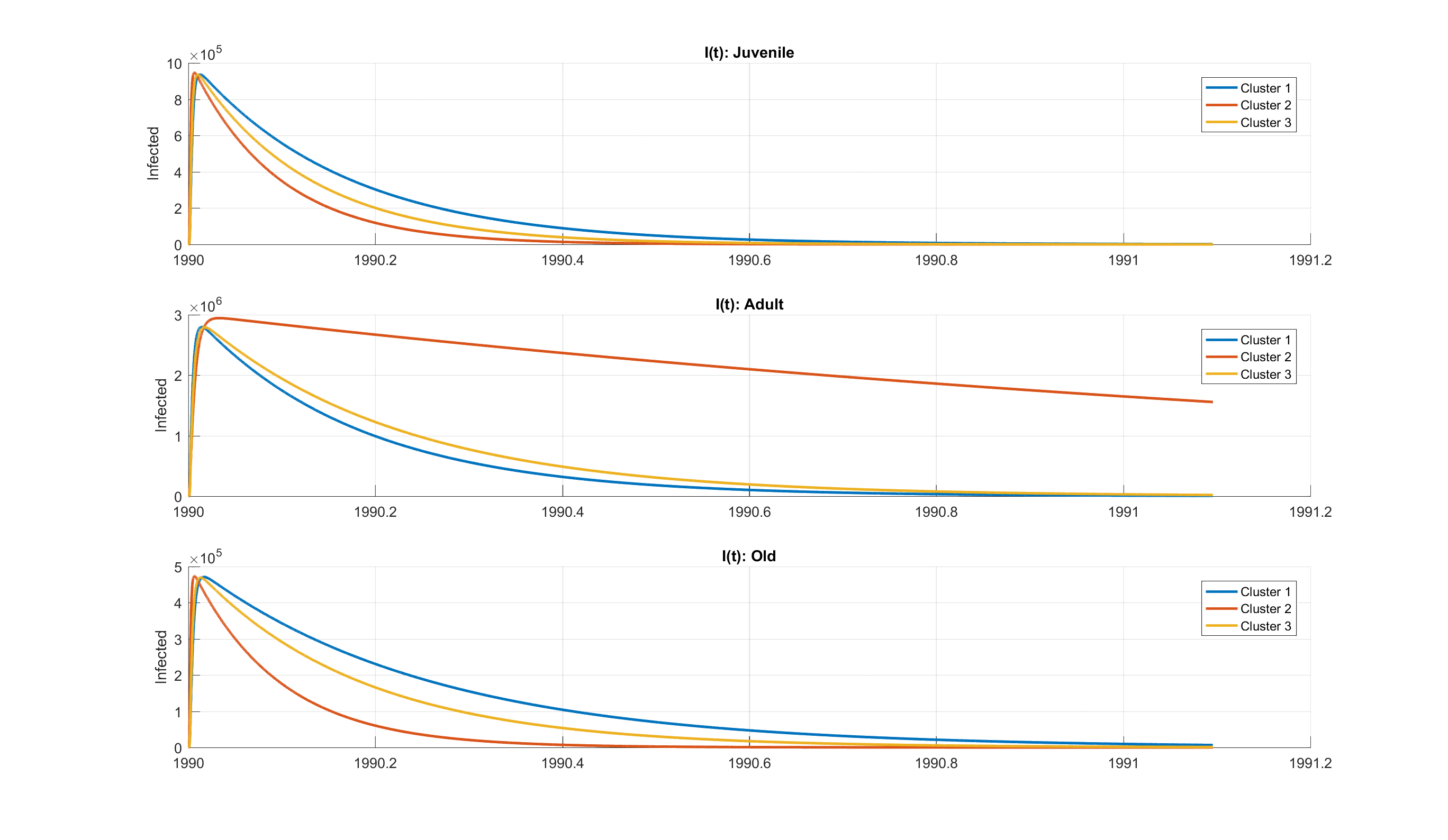}
        \subcaption{}
    \end{minipage}
     
    \vspace{0.3cm}

\centering
    \begin{minipage}{0.48\textwidth}
        \centering
        \includegraphics[width=\textwidth,height=6.5cm]{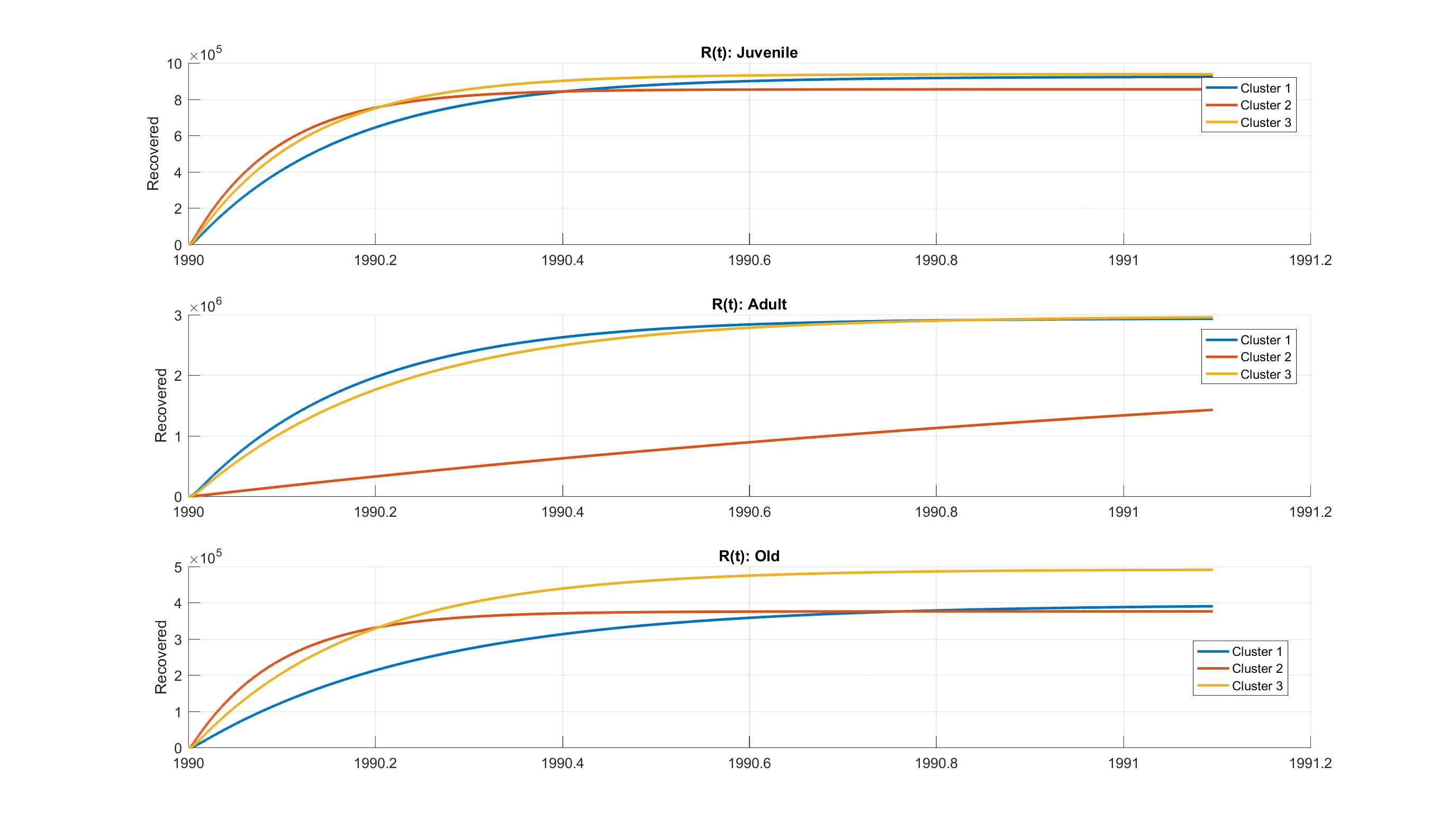}
        \subcaption{}
    \end{minipage}
    \hfill
    \begin{minipage}{0.48\textwidth}
        \centering
        \includegraphics[width=\textwidth,height=6.5cm]{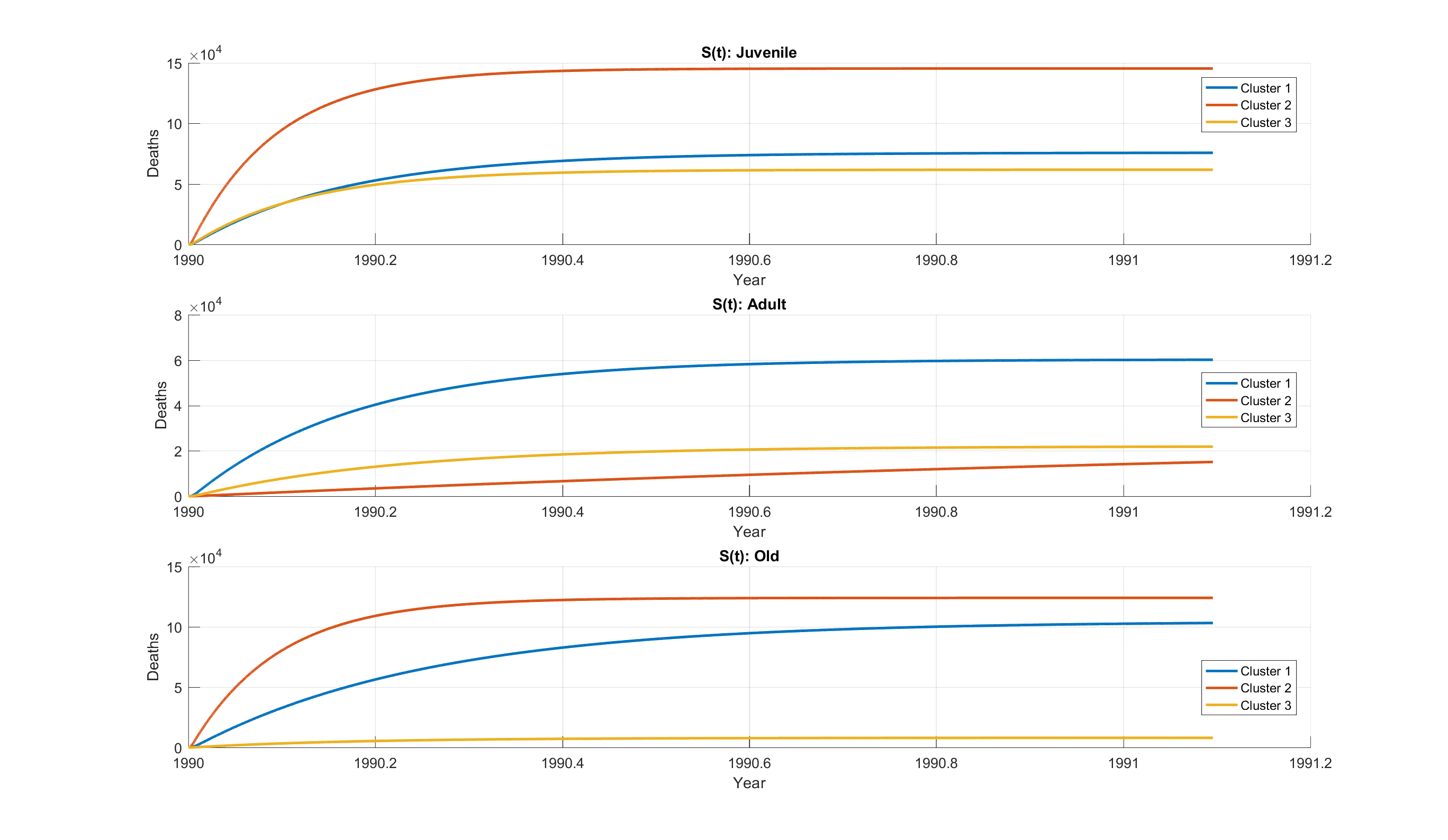}
        \subcaption{}
    \end{minipage}
     \caption{(a) Susceptible trajectories $S(t)$, (b) Infected trajectories $I(t)$, (c) Recovered trajectories $R(t)$ and (d) Death trajectories $D(t)$ for juveniles, adults, and the elderly across the three clusters.}
     \label{fig:cluster_si}
\end{figure}

In contrast, Cluster~2 exhibits a high-transmission regime (Figure~\ref{fig:trajectories}(b)). Very strong transmission leads to rapid depletion of susceptibles in juveniles and the elderly, accompanied by sharp and early infection peaks in these age groups. Among adults, however, the infected population remains elevated for a longer duration, indicating sustained transmission despite slower susceptible depletion. The recovered class grows rapidly for juveniles and the elderly, whereas adult recoveries accumulate more gradually. Mortality in Cluster~2 is strongly age-dependent, with high cumulative deaths among juveniles and the elderly but comparatively low mortality in adults due to a small disease-induced adult mortality rate.

Cluster~3 represents an intermediate regime between these two extremes (Figure~\ref{fig:trajectories}(c)). Infection peaks are moderate and decay smoothly over time, with susceptible depletion occurring faster than in Cluster~1 but more slowly than in Cluster~2. Stronger saturation effects reduce effective transmission at high prevalence, preventing runaway growth while still allowing substantial spread. Mortality remains consistently lower than in Clusters~1 and~2, particularly among the elderly.

Age-stratified trajectories provide additional insight into these inter-cluster differences. The susceptible curves in Figure~\ref{fig:cluster_si}(a) reveal that juveniles in Cluster~2 experience the fastest depletion of susceptibles, followed by Cluster~3, while Cluster~1 maintains a larger susceptible pool for a longer duration. Among adults, Cluster~1 shows the most rapid decline in susceptibility, whereas Cluster~2 exhibits a comparatively slower depletion. In the elderly, the patterns again resemble those of juveniles, with Cluster~2 showing the steepest decline and Cluster~1 preserving the largest remaining susceptible fraction.

The infected curves in Figure~\ref{fig:cluster_si}(b) highlight age-dependent differences in epidemic intensity. Among juveniles, Cluster~2 produces the highest and earliest infection peak, followed by Cluster~3, while Cluster~1 shows a slower decay of infections. In adults, Cluster~2 dominates the infection dynamics, with the infected population remaining high over an extended period, whereas Clusters~1 and~3 exhibit more rapid post-peak declines. Among the elderly, infection patterns again mirror those of juveniles, with Cluster~2 leading to the steepest peak and Cluster~1 showing the slowest decay.

The recovered curves in Figure~\ref{fig:cluster_si}(c) follow similar trends. Cluster~2 generates the largest and fastest accumulation of recoveries among juveniles and the elderly, while Cluster~3 shows intermediate behavior and Cluster~1 accumulates recoveries more gradually. In adults, however, Clusters~1 and~3 result in larger recovered populations than Cluster~2, reflecting prolonged infection persistence and slower recovery in the latter.

The death trajectories in Figure~\ref{fig:cluster_si}(d) are shaped by the interplay between transmission intensity and age-specific mortality. For juveniles, Cluster~2 leads to the highest cumulative deaths, followed by Cluster~1, while Cluster~3 remains lower. Among adults, Cluster~1 produces the highest mortality, whereas Cluster~2 yields the lowest despite its high infection burden. In the elderly, Cluster~2 exhibits the steepest and largest death curve, Cluster~1 shows substantial but lower mortality, and Cluster~3 remains significantly lower due to reduced elderly mortality rates.

Overall, these results underscore the critical role of parameter heterogeneity, age structure, and incidence saturation in shaping epidemic outcomes, even when the underlying model structure remains unchanged.

\section{Simulation Study}

To evaluate the performance of the proposed age-structured SIRD model and the cluster recovery procedure, we conducted a simulation study under controlled conditions. The objectives were to assess whether the model could (i) recover the true clusters of states based on incidence, prevalence, and death trajectories, and (ii) accurately estimate the SIRD parameters ($\beta$, $\gamma$, $\mu$) across age groups. The analysis considers $T=32$ years of data from $n_{\text{states}}=10$ states, clustered into $n_{\text{clusters}}=3$ groups, with age stratification into Adult, Juvenile, and Old categories. True SIRD parameters for each cluster and age group were chosen to reflect realistic epidemiological variation (Table~\ref{tab:true_params}).

\begin{table}[h!]
\centering
\begin{tabular}{lcccc}
\hline
Cluster & Age Group & $\beta$ & $\gamma$ & $\mu$ \\
\hline
1 & Adult    & 6.5  & 5.5 & 0.1 \\
1 & Juvenile & 7.5  & 5.5 & 0.5 \\
1 & Old      & 5.5  & 3.1 & 0.8 \\
2 & Adult    & 2.4  & 0.6 & 0.005 \\
2 & Juvenile & 13  & 9.1 & 1.5 \\
2 & Old      & 13 & 8.05 & 2.6 \\
3 & Adult    & 4.5  & 2.9 & 0.03 \\
3 & Juvenile & 8.9  & 7.7 & 0.5 \\
3 & Old      & 6.6  & 5.5 & 0.09 \\
\hline
\end{tabular}
\caption{True SIRD parameters used for simulation.}
\label{tab:true_params}
\end{table}

\noindent The cluster-based SIRD compartments were simulated using discrete-time updates:
\[
\begin{aligned}
s_{k,t+1} &= \max(s_{k,t} - \beta_k s_{k,t} i_{k,t}, 0), \\
i_{k,t+1} &= \max(i_{k,t} + \beta_k s_{k,t} i_{k,t} - \gamma_k i_{k,t} - \mu_k i_{k,t}, 0), \\
r_{k,t+1} &= r_{k,t} + \gamma_k i_{k,t}, \\
d_{k,t+1} &= d_{k,t} + \mu_k i_{k,t}.
\end{aligned}
\]

\noindent Observed data were generated by adding realistic measurement noise:
\[
\begin{aligned}
y^\text{inc}_{k,t} &\sim \text{Poisson}(\text{Inc}_{k,t} \cdot 10^5), \\
y^\text{prev}_{k,t} &\sim \text{Normal}(\text{Prev}_{k,t}, 0.01^2), \\
y^\text{death}_{k,t} &\sim \text{Poisson}(\text{Death}_{k,t}).
\end{aligned}
\]

\noindent Cluster membership for the 10 states was randomly assigned, ensuring all three age groups were represented in each state.

\subsection{Cluster Recovery Assessment}

State-level features were computed as the mean incidence, prevalence, and death over time. K-means clustering was applied to these features to estimate clusters. Cluster recovery was evaluated using the Adjusted Rand Index (ARI):
\[
\text{ARI} = \frac{\text{Index} - \text{Expected Index}}{\text{Max Index} - \text{Expected Index}},
\]
where ARI = 1 indicates perfect recovery, ARI = 0 indicates random clustering, and negative values indicate worse-than-random recovery.

\noindent From the simulation study, the true and estimated clusters were:

\begin{verbatim}
True labels:      3, 3, 3, 2, 3, 2, 2, 2, 3, 1
Estimated labels: 3, 3, 3, 2, 3, 2, 2, 2, 1, 3
\end{verbatim}

\noindent The contingency table between true and estimated clusters:

\begin{center}
\begin{tabular}{c|ccc}
\hline
True / Estimated & 1 & 2 & 3 \\
\hline
1 & 0 & 0 & 1 \\
2 & 0 & 4 & 0 \\
3 & 1 & 0 & 4 \\
\hline
\end{tabular}
\label{tab:ari}
\end{center}

The resulting ARI was 0.612, indicating strong agreement between true and estimated cluster assignments.

\subsection{Parameter Recovery}

Posterior mean estimates of SIRD parameters closely matched the true values (Table~\ref{tab:comp}), confirming that the Bayesian estimation framework accurately recovers age-specific epi-

\begin{table}[h!]
\centering
\small
\begin{tabular}{llcccccc}
\hline
\textbf{Cluster} & \textbf{Age Group} &
\textbf{True $\beta$} & \textbf{Est. $\beta$} &
\textbf{True $\gamma$} & \textbf{Est. $\gamma$} &
\textbf{True $\mu$} & \textbf{Est. $\mu$} \\
\hline

1 & Adult    & 6.5 & 6.410 & 5.5 & 4.778 & 0.1 & 0.247 \\
1 & Juvenile & 7.5 & 6.275 & 5.5 & 5.921 & 0.5 & 0.380 \\
1 & Old      & 5.5 & 4.216 & 3.1 & 4.159 & 0.8 & 0.524 \\

2 & Adult    & 2.4 & 4.456 & 0.6 & 0.571 & 0.005 & 0.0023 \\
2 & Juvenile & 13 & 12.882 & 9.1 & 8.780 & 1.5 & 0.86 \\
2 & Old      & 13 & 15.950 & 8.05 & 5.220 & 2.6 & 1.057 \\

3 & Adult    & 4.5 & 3.236  & 2.9 & 3.859 & 0.03 & 0.0841 \\
3 & Juvenile & 8.9 & 6.908 & 7.7 & 7.927 & 0.5 & 0.102 \\
3 & Old      & 6.6 & 8.170  & 5.5 & 3.053  & 0.09 & 0.0443 \\

\hline
\end{tabular}
\caption{Comparison of true vs. estimated SIRD parameters ($\beta$, $\gamma$, $\mu$) for each cluster and age group.}
\label{tab:comp}
\end{table}

\noindent demic dynamics. These results confirm that the proposed age-structured SIRD model and the clustering procedure can reliably recover underlying heterogeneity in epidemic dynamics, providing confidence for application to the real GBD chickenpox data.

\section{Discussion}
In this paper, we developed an age-structured Bayesian SIRD model with a saturating Holling-type incidence function to study the long-term transmission of chickenpox in India from 1990 to 2021. The model jointly analyzes incidence, prevalence, and mortality data from the Global Burden of Disease database, which allows us to better estimate transmission, recovery, and death rates while accounting for uncertainty and underreporting. By combining age stratification with clustering of states, the model captures important differences in disease dynamics across age groups and regions.

The results show clear and consistent epidemiological patterns. Across all clusters, juveniles are the main drivers of transmission, likely due to higher contact rates and susceptibility. Adults generally experience moderate transmission with low mortality, while older individuals face a much higher risk of death because of slower recovery and higher disease-related mortality. The cluster-based analysis highlights strong regional heterogeneity, indicating that chickenpox transmission in India cannot be adequately described by a single national-level model. These findings suggest that age-targeted public health strategies, such as vaccination and monitoring among children, along with protective measures for the elderly, are especially important.

From a methodological point of view, the Holling-type incidence function allows transmission to saturate at high infection levels, which avoids unrealistically rapid epidemic growth while keeping the model interpretable. The Bayesian framework explicitly accounts for measurement error and reporting bias, leading to more reliable parameter estimates. The simulation study further confirms that the proposed approach can recover both clusters and age-specific model parameters with good accuracy.

Several limitations of the study should be noted. First, the analysis is based on annual data, which limits the ability to capture short-term outbreak dynamics. For this reason, we used a discrete-time approximation of the SIRD model that is suitable for long-term trend analysis but not for short-term forecasting. Second, parameter identifiability remains a challenge in compartmental models, especially when transmission rates and reporting factors are closely linked. Although using multiple data sources helps reduce this problem, some uncertainty remains. 
Finally, clustering was performed as a post-hoc step using K-means, which is simple and interpretable but not fully integrated into the Bayesian model.

Despite these limitations, the proposed framework provides a flexible and reproducible approach for studying long-term infectious disease dynamics in heterogeneous populations. Future work could extend the model by jointly estimating clusters within a fully Bayesian framework, incorporating vaccination coverage, or accounting for spatial dependence between neighboring states. Overall, this study shows that combining age-structured epidemic models, Bayesian inference, and clustering methods can provide useful insights into the uneven spread of chickenpox in India and inform targeted public health strategies.

\section*{Acknowledgments including declarations}
The authors would like to thank all contributors and collaborators 
who provided valuable insights and support during the preparation of this work. 

\vspace{0.1 in}

\noindent \textbf{Statements of Ethical Approval:}
This study does not involve any human participants or animal experiments. 
Hence, formal ethical approval was not required. 

\vspace{0.1 in}

\noindent \textbf{Funding:}
This research did not receive any specific grant from funding agencies 
in the public, commercial, or not-for-profit sectors. 

\vspace{0.1 in}

\noindent \textbf{Competing Interests:}
The authors declare that they have no known competing financial interests 
or personal relationships that could have appeared to influence the 
work reported in this paper. 

\vspace{0.1 in}

\noindent \textbf{Data Availability:}
All data generated or analyzed during this study are included in this article. 
Additional datasets are available from the corresponding author upon request.

\end{document}